\documentclass[a4paper,fleqn]{cas-dc}

\usepackage[numbers]{natbib}

\usepackage{xurl}
\usepackage{cleveref}
\newcommand{\cvec}[1]{\vec{#1}}
\def\tsc#1{\csdef{#1}{\textsc{\lowercase{#1}}\xspace}}
\tsc{WGM}
\tsc{QE}

\begin{document}
\let\WriteBookmarks\relax
\def\floatpagepagefraction{1}
\def\textpagefraction{.001}

\shorttitle{Rapid Vacuum Response Calculations for Non-axisymmetric Plasma Geometries}    

\shortauthors{Halpern, Logan, N\"uhrenberg, Paul, and Paz-Soldan}  

\title [mode = title]{Rapid Vacuum Response Calculations for Non-axisymmetric Plasma Geometries}  

%

\author[1]{Jacob Halpern}[orcid=0000-0001-9370-8160]

\cormark[1]

\ead{jmh2363@columbia.edu}

\credit{Methodology, Software, Investigation, Writing - Original Draft, Writing - Review $\&$ Editing, Funding acquisition}

\affiliation[1]{organization={Columbia University},
            city={New York},
            state={NY},
            country={United States}}

\author[1]{Nikolas Logan}[orcid=0000-0002-3268-7359]
\credit{Conceptualization, Software, Supervision, Writing - Review $\&$ Editing}

\author[2]{Carolin N\"uhrenberg}[orcid=0000-0002-3568-2035]
\credit{Investigation, Writing - Review $\&$ Editing}

\author[1]{Elizabeth Paul}[orcid=0000-0002-9355-5595]
\credit{Conceptualization, Supervision, Writing - Review $\&$ Editing}

\author[1]{Carlos Paz-Soldan}[orcid=0000-0001-5069-4934]
\credit{Conceptualization, Funding acquisition, Supervision, Writing - Review $\&$ Editing}

\affiliation[2]{organization={Max-Planck-Institut f\"ur Plasmaphysik},
            city={Greifswald},
            country={Germany}}

\cortext[1]{Corresponding author}

\begin{abstract}
We present a fast, open-source calculation of the vacuum response matrix and surface inductance in non-axisymmetric geometries capable of running with up to $10^5$ boundary points in tens of seconds on a laptop. Ideal magneto-hydrodynamic (MHD) stability modeling for free boundary modes requires the perturbed vacuum energy. We extend the VACUUM code to non-axisymmetric geometries in the new Julia implementation of the Generalized Perturbed Equilibrium Code, combining a collocation approach to the boundary integral equations with a high-order singularity correction scheme. The correction weights are assembled into the dense boundary operators of the collocation formulation, including a conducting wall, and the structure of these operators under field-period and stellarator symmetry is exploited to reduce storage and solve cost. We find that the vacuum response matrix converges at fourth order or greater in the grid spacing, reaching a given accuracy on a coarser grid than previous implementations without the high-order method. The non-axisymmetric calculation converges to the axisymmetric VACUUM result for a tokamak and the CAS3D result for a W7-X stellarator equilibrium. This work enables fast free boundary ideal MHD stability and perturbed equilibrium calculations in stellarators.
\end{abstract}



\begin{keywords}
 ideal MHD \sep boundary integral equations \sep fusion energy \sep plasma \sep stellarators
\end{keywords}

\maketitle

\begin{center}\rule{\columnwidth}{0.4pt}\end{center}

\section{Introduction}

Magnetic confinement is one of the most promising routes to clean energy from nuclear fusion, in which magnetic fields hold a hot, dense plasma away from material surfaces for long enough that fusion reactions release more energy than is invested in sustaining them. However, the same confinement produces steep gradients within the plasma volume, supplying the free energy for instabilities capable of rapidly degrading confinement. Predicting and optimizing the plasma equilibrium for stability is therefore essential for bringing fusion power to the grid.

The stability of a magnetically confined plasma is determined through the ideal magneto-hydrodynamic (MHD) energy principle, which states that an equilibrium is unstable if there exists a mode with plasma displacement $\boldsymbol{\xi}$ that makes the perturbed potential energy $\delta W$ negative \cite{bernstein1958}. Perturbations which displace the equilibrium boundary, called free boundary modes, require evaluation of both the plasma and vacuum components of the perturbed energy, $\delta W = \delta W_P + \delta W_V$ \cite{merkel1996,glasser2016}. The perturbed vacuum energy solution can also be employed to calculate vacuum fields inside or outside the plasma \cite{merkel1986}, eddy current patterns induced in nearby conducting walls \cite{chance1997}, and the surface inductance matrix required for obtaining perturbed plasma equilibria from externally applied fields \cite{atanasiu1999, boozer2001, maslovsky2005, boozerPerturbedPlasmaEquilibria2006, parkComputationThreedimensionalTokamak2007a, nuhrenberg2012}.

In the axisymmetric geometries of tokamaks, MHD stability calculations are quite mature, with many codes developed and cross-validated over decades of research \cite{grimm1976,grimm1983,gruber1981,bernard1981,cheng1987}. The DCON code \cite{glasser2016} determines fixed boundary tokamak stability through a variational approach, reducing the plasma response to a matrix $\mathbb{W}_P$ at the boundary, which can be combined with the vacuum response matrix $\mathbb{W}_V$ from the VACUUM code \cite{chance1997,chance2007}. The eigenvalues of this combined response matrix then give a fast determination of free boundary stability without the full eigenmode calculation. For stable plasmas, the VACUUM solution also provides the surface inductance matrix to construct perturbed equilibria in the Generalized Perturbed Equilibrium Code (GPEC) \cite{parkComputationThreedimensionalTokamak2007a,park2017}.

Linear MHD stability modeling in the non-axisymmetric geometries of stellarators relies on comparatively few codes \cite{anderson1990,nuhrenberg2021}, each of which solves the full eigenvalue problem including the plasma kinetic energy. The DCON code has recently been extended to stellarators with stepped-pressure profiles \cite{glasser2020} and nested flux surfaces \cite{glasser2025}, allowing fast calculation of the non-axisymmetric $\delta W_P$. However, the corresponding non-axisymmetric $\delta W_V$ calculation has no general, openly available implementation, as existing calculations \cite{merkel1986,hirshman1986,merkel1996,merkel2015,strumberger2016,weyens2019,strumberger2023} are not openly distributed, are coupled to a particular stability code, or are not formulated in the $\mathbb{W}_V$ matrix formalism required by DCON.

The perturbed vacuum energy is generally obtained by recasting the volume integral of the perturbed magnetic energy in the vacuum region as a boundary integral equation over the plasma-vacuum boundary using Green's functions. While this reduces the dimensionality of the problem, it introduces singularities that invalidate standard numerical quadrature and require specialized treatment. In general, existing solvers subtract the leading-order term of the singularity and integrate it analytically, and primarily differ in the use of Fourier \cite{merkel1986,hirshman1986,merkel2015} versus collocation \cite{chance1997} methods. Higher-order singularity correction schemes have since been developed for boundary integral equations \cite{bruno2001,bruno2001a,ying2006} and recently applied to plasma physics \cite{malhotra2019,malhotra2019a}.

In this work, we present the extension of the VACUUM code to non-axisymmetric geometries, built as part of the new, open-source Julia version of GPEC \cite{logan2026} to be released as version 2.0 and available at \url{https://github.com/OpenFUSIONToolkit/GPEC}. We have combined the collocation approach of the original VACUUM code with the high-order singularity correction scheme of \citet{malhotra2019, malhotra2019a} to enable fast and accurate vacuum response calculations. We derive the non-axisymmetric vacuum response matrix and surface inductance matrix in \cref{sec:response_matrix}, where we show that both reduce to the same pair of discretized layer potential operators. Because the kernels of these operators are singular, the boundary quadrature requires the singularity correction scheme noted above, which we describe and adapt to a collocation matrix formulation with a conducting wall in \cref{sec:singular}. In \cref{sec:field-period-symmetry} we show that field-period symmetry makes these same operators block-circulant, reducing storage and solve cost, and in \cref{sec:implementation-results} we discuss our numerical implementation and benchmark against the axisymmetric VACUUM code and the CAS3D stellarator code.

\section{Non-axisymmetric Vacuum Response Matrix} \label{sec:response_matrix}

In the DCON formalism \cite{glasser2016}, the free boundary stability problem is solved by constructing the total response matrix from its plasma and vacuum components,
\begin{equation}
    \label{eq:matrix_formalism}
    \delta W = \delta W_P + \delta W_V = \cvec{\Xi}^{\dagger} \left(\mathbb{W}_P + \mathbb{W}_V\right) \cvec{\Xi},
\end{equation}
where $\cvec{\Xi}$ is a vector of normal displacement amplitudes on the plasma boundary spanning a given mode space and $\dagger$ denotes the adjoint. Throughout, bold symbols such as $\boldsymbol{\xi}$ denote vectors in physical space, overarrows such as $\cvec{\Xi}$ denote column vectors of discrete values (Fourier coefficients in the mode basis or values at the collocation points introduced below), and blackboard-bold symbols such as $\mathbb{W}_P$ denote matrices. If all eigenvalues of the total response matrix are non-negative, the equilibrium is stable to the set of modes. DCON readily calculates the plasma response matrix $\mathbb{W}_P$; we focus on the vacuum response $\mathbb{W}_V$ in this work.

To obtain the vacuum response matrix, we start with the expression for the perturbed vacuum energy,
\begin{equation}
\delta W_V = \frac{1}{2\mu_0}\int |\delta \mathbf{B}|^2 \, dV,
\label{eq:dWv_volume}
\end{equation}
where $dV$ is the volume element, $\delta \mathbf{B}$ is the perturbed field, and the integral is taken over the vacuum region. The vacuum region is current-free, so $\nabla\times\delta \mathbf{B} = 0$ there and the perturbed field is the gradient of a scalar potential $\varphi$,
\begin{equation}
\label{eq:B1_grad_phi}
\delta \mathbf{B} = \nabla\varphi.
\end{equation}

The boundary $\Gamma$ of the vacuum region consists of the plasma surface $\Gamma_p$ and (optionally) a conducting wall $\Gamma_w$. Each is described by a smooth, doubly periodic parameterization of the Cartesian position $\mathbf{r}(\theta,\zeta)$ with poloidal angle $\theta$ and toroidal angle $\zeta$, both $\in [0,2\pi)$.
Combining \cref{eq:dWv_volume} and \cref{eq:B1_grad_phi} with the divergence theorem, the volume integral becomes a surface integral over the boundary $\Gamma$ of the vacuum region,
\begin{equation}
\label{eq:dWv_surface}
\begin{split}
\delta W_V &= \frac{1}{2\mu_0}\int_{\Gamma_p} \varphi(\mathbf{r})\,\nabla\varphi^*(\mathbf{r})\cdot d\mathbf{S} \\
&= -\frac{1}{2\mu_0}\int_{\Gamma_p} \varphi(\mathbf{r})\,b^*(\mathbf{r})\,d\theta\,d\zeta,
\end{split}
\end{equation}
where we have introduced the normal perturbed field on the plasma boundary,
\begin{equation}
    \label{eq:b_def}
    b \equiv \mathit{J}\,\delta \mathbf{B}\cdot\nabla\psi,
\end{equation}
in the form supplied by DCON, where $\psi \in [0,1]$ is the normalized flux-surface label of the plasma equilibrium, $\Gamma_p$ its bounding surface $\psi = 1$, and $\mathit{J}$ the Jacobian of its flux coordinates. The boundary integral operators are oriented by $\mathbf{n} = \sigma\,\partial_\theta\mathbf{r}\times\partial_\zeta\mathbf{r}$, the unnormalized surface normal with $\sigma = \pm 1$ chosen on each surface so that $\mathbf{n}$ points out of the vacuum region. Leaving it unnormalized carries the area weighting $dS = |\mathbf{n}|\,d\theta\,d\zeta$ into the integrals below. On $\Gamma_p$, $\mathbf{n}$ therefore points into the plasma, antiparallel to $\nabla\psi$, and the identity $\partial_\theta\mathbf{r}\times\partial_\zeta\mathbf{r} = \mathit{J}\nabla\psi$ sets $\sigma = -\mathrm{sgn}\,\mathit{J}$. For $\mathit{J} > 0$,
\begin{equation}
    \label{eq:b_normal_relation}
    \delta \mathbf{B}\cdot\mathbf{n} = \nabla\varphi\cdot\mathbf{n} = -b ,
\end{equation}
which is the origin of the sign in \cref{eq:dWv_surface}. While the surface integral is over the entire vacuum boundary, only the plasma surface contributes because $b = 0$ on $\Gamma_w$.

\Cref{eq:dWv_surface} then requires two independent boundary quantities: the potential $\varphi$, and the normal perturbed field $b$, the latter being the Neumann data supplied by the plasma solution. Using Green's second identity and $\nabla^2 \varphi = 0$, we can derive a Fredholm integral equation of the second kind relating these two quantities on the boundary,
\begin{equation}
    \label{eq:fredholm}
    \begin{split}
    \frac{1}{2}\varphi(\mathbf{r}) &+ \int_\Gamma \varphi(\mathbf{r}')\,K(\mathbf{r},\mathbf{r}')\, d\theta' d\zeta' \\
    &= -\int_{\Gamma_p} G(\mathbf{r},\mathbf{r}')\,b(\mathbf{r}')\, d\theta' d\zeta' ,
    \end{split}
\end{equation}
where a prime denotes evaluation at the source point, $G(\mathbf{r},\mathbf{r}') = \left(4\pi|\mathbf{r} - \mathbf{r}'|\right)^{-1}$ is the free space Green's function, and
\begin{equation}
    \label{eq:double_layer_kernel}
    K(\mathbf{r},\mathbf{r}') \equiv \nabla' G(\mathbf{r},\mathbf{r}') \cdot \mathbf{n}'.
\end{equation}
The two integrals in \cref{eq:fredholm} are the double- and single-layer potentials of classical potential theory with double-layer kernel $K$ and single-layer kernel $G$, respectively. The double-layer potential is discontinuous across the surface, and for a smooth boundary its limits from either side of $\Gamma$ differ in \cref{eq:fredholm} by $\pm\varphi/2$. \Cref{eq:fredholm} is the limit taken from the vacuum side; the plasma-side limit carries the opposite sign and is used in \cref{sec:surface_inductance}. As in \cref{eq:dWv_surface}, the single-layer integral is restricted to $\Gamma_p$ because $b = 0$ on $\Gamma_w$. The double-layer integral spans all of $\Gamma$ because $\varphi$ is nonzero on $\Gamma_w$ unless the wall is infinitely far away. These two operators appear throughout the rest of this work, as their discrete form provides both $\mathbb{W}_V$ below and the surface inductance matrix in \cref{sec:surface_inductance}, while \cref{sec:singular} discusses how to treat their singularities.
The solution to \cref{eq:fredholm} can be written in terms of a response function $R(\mathbf{r},\mathbf{r}')$,
\begin{equation}
    \label{eq:response_function}
    \varphi(\mathbf{r}) = -\int_{\Gamma_p} R(\mathbf{r},\mathbf{r}')\,b(\mathbf{r}')\, d\theta' d\zeta'.
\end{equation}
The observer point $\mathbf{r}$ ranges over the full boundary $\Gamma$ while the source point $\mathbf{r}'$ is confined to $\Gamma_p$, an asymmetry inherited directly by the discrete operators below.
Substituting \cref{eq:response_function} into \cref{eq:dWv_surface}, we obtain
\begin{equation}
    \label{eq:dWv_surface_response_function}
    \delta W_V = \frac{1}{2\mu_0}\int_{\Gamma_p} d\theta\, d\zeta \int_{\Gamma_p} d\theta' d\zeta'\, b^*(\mathbf{r})\, R(\mathbf{r},\mathbf{r}')\,b(\mathbf{r}').
\end{equation}

We now discretize \cref{eq:dWv_surface_response_function} to obtain the form desired in \cref{eq:matrix_formalism}. Following \citet{chance1997}, we solve by collocation with $N = N_\theta N_\zeta$ collocation points on both the plasma and wall surfaces, equally spaced in $\theta$ and $\zeta$ on the plasma surface and in a separate parameterization on the wall. The discrete form of \cref{eq:fredholm} for the $i$th observer and $j$th source point can then be written as
\begin{equation}
    \label{eq:discrete_fredholm}
    \begin{split}
    \frac{1}{2} \varphi(\mathbf{r}_i) &+ \frac{4\pi^2}{N}\sum_{j=1}^{2N} w_{ij} K(\mathbf{r}_i,\mathbf{r}_j) \varphi(\mathbf{r}_j) \\
    &= -\frac{4\pi^2}{N} \sum_{j=1}^{N} w_{ij} G(\mathbf{r}_i,\mathbf{r}_j) b(\mathbf{r}_j),
    \end{split}
\end{equation}
where $w_{ij}$ are the integral discretization weights and we have substituted $d\theta\,d\zeta = 4\pi^2/N$ due to the uniform grid. The first $N$ collocation points are on the plasma surface and the second $N$ on the wall surface. The weights here are nontrivial due to the singularity that occurs in the Green's function when $\mathbf{r}_i = \mathbf{r}_j$. In matrix form, this becomes
\begin{equation}
    \label{eq:discrete_fredholm_matrix}
    \left(\frac{1}{2}\mathbb{I}_{2N} + \mathbb{D} \right) \cvec{\varphi} = -\mathbb{S} \cvec{b},
\end{equation}
where $\cvec{\varphi}$ and $\cvec{b}$ are column vectors of length $2N$ and $N$ of $\varphi$ and $b$ at the collocation points, $\mathbb{D}$ is the $2N \times 2N$ double-layer potential operator, and $\mathbb{S}$ is the $2N \times N$ single-layer potential operator. When the conducting wall is absent, or equivalently assumed infinitely far away, there are only $N$ collocation points on the plasma surface and $\mathbb{D}$, $\mathbb{S}$, and the identity are all $N \times N$ matrices. The solution of \cref{eq:discrete_fredholm_matrix} then gives the discrete form of the response function from \cref{eq:response_function},
\begin{equation}
    \label{eq:discrete_response_function}
    \mathbb{R} = \left(\frac{1}{2}\mathbb{I}_{2N} + \mathbb{D} \right)^{-1} \mathbb{S},
\end{equation}
such that $\cvec{\varphi} = -\mathbb{R}\cvec{b}$.

The perturbed vacuum energy is then given by
\begin{equation}
    \label{eq:dWv_discrete}
    \delta W_V = \frac{1}{2\mu_0} \frac{4\pi^2}{N} \sum_{i=1}^N w_i\, b^*(\mathbf{r}_i) \left[\mathbb{R}_p \cvec{b} \right]_i = \frac{1}{2\mu_0} \frac{4\pi^2}{N} \cvec{b}^\dagger \mathbb{R}_p \cvec{b},
\end{equation}
where we have used periodic trapezoidal weights $w_i = 1$. Here $\mathbb{R}_p$ denotes the first $N$ rows of $\mathbb{R}$ corresponding to the magnetic potential on the plasma surface. 
To obtain the form desired in \cref{eq:matrix_formalism}, we first assume the linear perturbed quantities vary like $e^{i(m\theta - n\zeta)}$ to project the length $N$ vectors over the collocation points onto our mode basis with $M_\theta$ and $M_\zeta$ modes retained in the poloidal and toroidal directions, respectively, for a total of $M = M_\theta M_\zeta$ modes. We now take $\theta$ and $\zeta$ on $\Gamma_p$ to be straight-field-line angles, chosen such that field lines are straight in the $(\theta,\zeta)$ plane. A derivative along the magnetic field is then $\mathit{J}\mathbf{B}\cdot\nabla = \chi'(\partial_\theta + q\,\partial_\zeta)$ where $\chi(\psi)$ is the poloidal flux function with $\chi' = d\chi/d\psi$ and $q(\psi)$ is the safety factor. This expression can also be written in terms of the rotational transform $\iota = 1/q$ more commonly used in stellarators; however, we use $q$ below for consistency with prior work \cite{chance1997,glasser2016,glasser2025}. Combined with the relation $\delta \mathbf{B} = \nabla \times (\boldsymbol{\xi} \times \mathbf{B} )$, we can then express $\cvec{b}$ in terms of the normal displacement amplitudes on the boundary
\begin{equation}
    \label{eq:ideal_MHD_relationship}
    \begin{split}
        b(\mathbf{r}) &= \sum_{m,n} \Phi_{mn} e^{i(m\theta - n\zeta)} \\
          &= i \chi' \sum_{m,n} (m - n q) \xi^\psi_{mn} e^{i(m\theta - n\zeta)},
    \end{split}
\end{equation}
where $\Phi_{mn}$ are the flux-like Fourier coefficients of the $mn$th mode of $b$ and $\xi^\psi_{mn}$ are the equivalent Fourier coefficients for $\boldsymbol{\xi}\cdot\nabla\psi$.
This operation can be written as a matrix multiplication
\begin{equation}
    \label{eq:fourier_projection}
    \cvec{b} = i \chi'\mathbb{F}  \mathbb{Q}  \cvec{\Xi},
\end{equation}
where $\mathbb{F}$ is the $N \times M$ Fourier projection matrix, $\mathbb{Q}$ is a diagonal $M \times M$ matrix with $m - n q$ for each mode, and $\cvec{\Xi}$ is the vector of $\xi^\psi_{mn}$ identical to that in \cref{eq:matrix_formalism}. The entries of $\mathbb{F}$ contain the Fourier basis functions $\exp(i(m\theta - n\zeta))$, with rows indexed by the collocation points $\mathbf{r}_i$ and columns by the mode pairs $(m,n)$. Substituting into \cref{eq:dWv_discrete} gives
\begin{equation}
    \label{eq:dWv_surface_response_function_fourier}
    \delta W_V = \frac{\chi'^2}{2\mu_0} \frac{4\pi^2}{N} \cvec{\Xi}^{\dagger} \mathbb{Q} \mathbb{F}^\dagger \mathbb{R}_p \mathbb{F} \mathbb{Q} \cvec{\Xi},
\end{equation}
where we have made use of $\mathbb{Q} = \mathbb{Q}^\dagger$.
Comparing to \cref{eq:matrix_formalism}, we see that the vacuum response matrix is given by
\begin{equation}
    \label{eq:Wv_fourier}
    \mathbb{W}_V = \frac{\chi'^2}{2\mu_0} \frac{4\pi^2}{N} \mathbb{Q} \mathbb{F}^\dagger \mathbb{R}_p \mathbb{F} \mathbb{Q}.
\end{equation}
This expression compresses the preceding derivation into a single matrix operation. The matrix $\mathbb{R}_p$ is a property of the boundary alone, giving the response of the magnetic scalar potential $\varphi$ produced by a normal field perturbation $b$, with $\mathbb{F}\mathbb{Q}$ converting the normal displacement amplitudes $\cvec{\Xi}$ into $b$ and its adjoint returning the energy to the mode space in which a fixed boundary code supplies $\mathbb{W}_P$. Only the quadrature weights inside $\mathbb{D}$ and $\mathbb{S}$ remain to be specified, which we derive in \cref{sec:singular}.

\subsection{Surface Inductance Matrix}
\label{sec:surface_inductance}

The same boundary integral operators $\mathbb{D}$ and $\mathbb{S}$ are also used to obtain the surface inductance required for computing perturbed equilibria, with no additional kernels required. We represent the plasma response by an equivalent surface current density on the plasma boundary,
\begin{equation}
    \label{eq:current_potential_def}
    \mathbf{j} = \hat{\mathbf{n}} \times \nabla \kappa ,
\end{equation}
in terms of a current potential $\kappa$ defined on $\Gamma_p$ with units of current, where $\hat{\mathbf{n}} = \mathbf{n}/|\mathbf{n}|$ retains the orientation defined above, pointing out of the vacuum region and into the plasma. The current potential can be expanded in our mode basis with coefficients
\begin{equation}
\label{eq:kappa}
    \kappa = \sum_{m,n} I_{mn} e^{i(m\theta - n\zeta)} .
\end{equation}
The surface inductance matrix $\mathbb{L}$ is a purely geometric quantity which supplies the linear relation between the perturbed surface current and the normal-field harmonics it produces at the boundary in the absence of a plasma,
\begin{equation}
    \label{eq:surf_inductance}
    \cvec{\Phi} = \mathbb{L} \cvec{I} ,
\end{equation}
where $\cvec{\Phi}$ is the vector of Fourier coefficients of $b\left(\mathbf{r}\right)$ with the relation $\cvec{b} = \mathbb{F}\cvec{\Phi}$, and $\cvec{I}$ is a vector of the coefficients $I_{mn}$ \cite{maslovsky2005, parkComputationThreedimensionalTokamak2007a}.

The perturbed vacuum surface current density can also be written in terms of the tangential jump in the perturbed vacuum field at the boundary,
\begin{equation}
    \mu_0 \mathbf{j} =\hat{\mathbf{n}} \times \delta \mathbf{B}^{\mathrm{in}} - \hat{\mathbf{n}} \times \delta \mathbf{B}^{\mathrm{out}},
\end{equation}
where the superscripts $\mathrm{in}$ and $\mathrm{out}$ denote the vacuum field evaluated just inside and outside the plasma surface, respectively. Using \cref{eq:B1_grad_phi}, this relation becomes
\begin{equation}
\label{eq:kappa_varphi_diff}
    \mu_0 \kappa = \varphi^{\mathrm{in}} - \varphi^{\mathrm{out}} .
\end{equation}
Taking the vacuum-side limit of \cref{eq:fredholm} has already given
\begin{equation}
    \cvec{\varphi}^{\mathrm{out}} = \mathbb{R}^{\mathrm{out}}\mathbb{F}\cvec{\Phi},
\end{equation}
where $\mathbb{R}^{\mathrm{out}} \equiv \mathbb{R}_p$, assuming there is no conducting wall present. The potential just inside the surface can be obtained by taking the plasma-side limit of \cref{eq:fredholm}, which reverses the sign of the jump term, such that
\begin{equation}
    \label{eq:fredholm_interior}
    \left(-\frac{1}{2}\mathbb{I}_N + \mathbb{D} \right) \cvec{\varphi}^{\mathrm{in}} = -\mathbb{S} \cvec{b},
\end{equation}
which defines the interior response matrix as
\begin{equation}
    \mathbb{R}^{\mathrm{in}} = \left(-\frac{1}{2}\mathbb{I}_N + \mathbb{D} \right)^{-1} \mathbb{S} .
\end{equation}
Inserting into \cref{eq:kappa_varphi_diff} and applying the Fourier operators along with $\mathbb{F}^\dagger\mathbb{F} = N\,\mathbb{I}_M$, we obtain
\begin{equation}
    \label{eq:current_projection}
    \cvec{I} = \frac{1}{\mu_0 N} \mathbb{F}^\dagger \left(\mathbb{R}^{\mathrm{out}} - \mathbb{R}^{\mathrm{in}}\right) \mathbb{F} \cvec{\Phi}.
\end{equation}
Comparing \cref{eq:current_projection} to \cref{eq:surf_inductance}, the surface inductance can be obtained from
\begin{equation}
    \label{eq:surf_inductance_matrix}
    \mathbb{L} = \mu_0 N \left[\mathbb{F}^\dagger \left(\mathbb{R}^{\mathrm{out}} - \mathbb{R}^{\mathrm{in}}\right) \mathbb{F}\right]^{-1} .
\end{equation}
Because the left-hand side of \cref{eq:fredholm_interior} differs from that of \cref{eq:discrete_fredholm_matrix} only in the sign of the identity term, the calculation of the boundary integral operators $\mathbb{D}$ and $\mathbb{S}$ need only be performed once, and the exterior and interior systems can then be formed and solved separately.

\section{Treatment of Green's Function Singularities}
\label{sec:singular}

The singularity in the Green's function as $\mathbf{r} - \mathbf{r}'\rightarrow 0$ means the trapezoidal rule and other Newton--Cotes formulae cannot supply the quadrature weights in \cref{eq:discrete_fredholm}. Rather than subtracting the leading-order singularity as in previous vacuum solvers \cite{chance1997, merkel1986}, we adapt the higher-order correction scheme of \citet{malhotra2019, malhotra2019a}, one of a family of high-order Nystr\"om corrections developed for boundary integral equations in scattering and potential theory \cite{bruno2001a, ying2006}. We present a self-contained description of the method in \cref{subsec:malhotra_quadrature} and adapt it to our matrix formulation in \cref{subsec:matrix_sing_quad}.

\Cref{fig:singular_quadrature} illustrates the complete scheme for a single observer point, with panels (a) and (b) corresponding to the quadrature construction of \cref{subsec:malhotra_quadrature} and panels (c) and (d) to its matrix form in \cref{subsec:matrix_sing_quad}.

\subsection{Description of the High-Order Quadrature Method}
\label{subsec:malhotra_quadrature}
\begin{figure*}[]
  \centering
  \includegraphics[width=\textwidth]{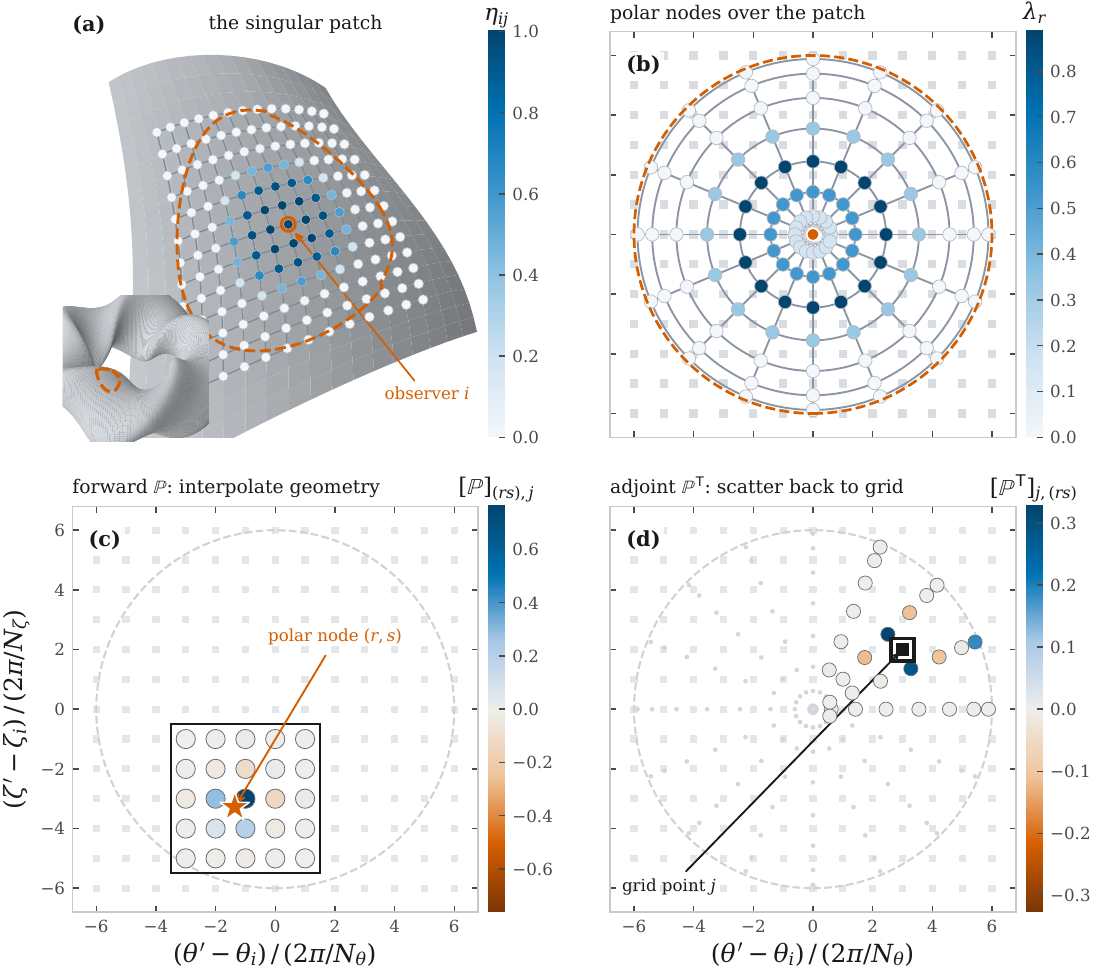}
  \caption{Singular quadrature scheme for a single observer point $i$ on a model rotating-ellipse boundary, shown for $R_\eta = 6$, $N_\rho \times N_\omega = 8 \times 16$, and $\ell = 5$ (reduced from the values used in \cref{sec:implementation-results} for legibility). (a) The $D_\eta \times D_\eta = 13 \times 13$ patch of collocation points centered on the observer, colored by $\eta_{ij}$, with the dashed orange curve marking $\rho = 1$; the inset shows the same patch in place on the full boundary. (b) The polar nodes in the $(\rho, \omega)$ grid, colored by the radial weight $\lambda_r$, with the underlying patch points as grey squares. (c) One row of $\mathbb{P}$, the $\ell^2 = 25$ patch points from which the marked polar node $(r,s)$ draws its geometry, colored by the interpolation weight. (d) The same matrix read by column, or equivalently one row of $\mathbb{P}^{\mathsf T}$: the $27$ polar nodes whose $\ell \times \ell$ stencil contains the marked grid point $j$ and which therefore scatter back onto it, with the remaining polar nodes shown as small grey dots. In (b) through (d) the axes are in grid cells from the observer.}
  \label{fig:singular_quadrature}
\end{figure*}

Both layer potentials in \cref{eq:fredholm} contain singular or sharply peaked kernels. For an observer point $\mathbf{r}_i$ and source surface $\Gamma_\ast \in \{\Gamma_p, \Gamma_w\}$, we write either potential in the common form
\begin{equation}
I^{H}_i[f;\Gamma_\ast] = \int_{\Gamma_\ast} H(\mathbf{r}_i,\mathbf{r}')f(\mathbf{r}') \,d\theta' d\zeta' ,
\label{eq:generic_layer_potential}
\end{equation}
where $H\in\{G,K\}$, with $(H,f)=(G,b)$ for the single-layer potential and $(H,f)=(K,\varphi)$ for the double-layer potential. We employ a partition of unity function $\eta_i = \eta(\mathbf{r}_i, \mathbf{r}')$ to split the integrals into a smooth  global integral and a local singular component,
\begin{equation}
I^{H}_i =\underbrace{\int_{\Gamma_\ast}(1-\eta_i)Hf\,d\theta' d\zeta'}_{I^{H}_{i,\mathrm{smooth}}} + \underbrace{\int_{\operatorname{supp}(\eta_i)}\eta_i Hf\,d\theta' d\zeta'}_{I^{H}_{i,\mathrm{sing}}}.
\label{eq:pou_split}
\end{equation}
The partition of unity is defined such that $\eta_i = 1$ around $\mathbf{r}_i$ and has compact support $\mathrm{supp}(\eta)$. We have found that the same definition of $\eta$ employed by \citet{malhotra2019} works well for our implementation,
\begin{equation}
    \eta(\mathbf{r}, \mathbf{r}') = \exp \left( -36 \rho^8 \right),
\end{equation}
where
\begin{equation}
    \label{eq:rho}
    \rho = \frac{1}{R_\eta} \sqrt{\left(  \frac{\theta - \theta'}{2\pi / N_{\theta}}\right)^2 + \left(\frac{\zeta - \zeta'}{2\pi / N_{\zeta}}\right)^2}.
\end{equation}
The integer $R_\eta$ sets the support of $\eta$, as $\eta$ falls to machine epsilon for $\rho > 1$ in double-precision arithmetic; it can be interpreted as the number of grid points away from the observer point in the poloidal and toroidal grid directions that receive the singular correction and have a nonzero value of $\eta$. The grid points with non-negligible $\eta$ then form a $D_\eta \times D_\eta$ grid of collocation points centered at the observer, where $D_\eta = 2R_\eta + 1$. The support of $\eta$ is a disk inscribed in this grid, shown in \cref{fig:singular_quadrature}(a) on a model non-axisymmetric boundary.

The smooth global integral can be computed using the periodic trapezoidal rule,
\begin{equation}
    \label{eq:smooth_discrete}
    I^H_{i,\mathrm{smooth}} \approx \frac{4\pi^2}{N} \sum_{j \in \Gamma_\ast} \left(1 - \eta_{ij} \right)H(\mathbf{r}_i,\mathbf{r}_j)f(\mathbf{r}_j).
\end{equation}
For the local singular integral, we perform a coordinate transformation into polar coordinates $\left( \rho, \omega\right)$,
\begin{equation}
    \theta - \theta' = \frac{2\pi}{N_\theta}R_\eta \rho \cos \omega, \qquad \zeta - \zeta' = \frac{2\pi}{N_\zeta}R_\eta \rho \sin \omega,
\end{equation}
where $\rho$ has precisely the same definition as in \cref{eq:rho}, and the singular integral can then be written as
\begin{equation}
    \label{eq:singular_integral}
    I^H_{i,\mathrm{sing}} = \frac{4\pi^2}{N} R_\eta^2 \int_0^{2\pi}\int_0^1 \eta(\rho) H(\mathbf{r}_i,\mathbf{r}') f(\mathbf{r}') \rho\,d\rho\,d\omega.
\end{equation}
The factor of $\rho$ from the transformation Jacobian cancels out the $1/\left| \mathbf{r} - \mathbf{r}'\right|$ in $G$, making the integral finite. It has been shown \cite{ying2006} that this approach extends to more strongly singular kernels, including double-layer kernels, so it applies to $K$. These integrals can then be computed to high accuracy using Gauss--Legendre integration in the radial direction and the trapezoidal rule in the angular direction, whose nodes are plotted in \cref{fig:singular_quadrature}(b). The rules differ by direction because the angular integrand is periodic, for which the trapezoidal rule already converges geometrically, while the radial integrand is not. Gauss--Legendre integration therefore restores geometric convergence in $\rho$ and avoids placing a quadrature node directly on the $\rho=0$ singularity.

The approach of \citet{malhotra2019a} does not directly discuss the inclusion of a conducting wall, which introduces integrals whose observer and source points lie on different surfaces. These integrals are non-singular, but can become near-singular when the conducting wall is close to the plasma. In our implementation, we apply the same partition-of-unity split to each observer-source pairing whether or not the integral is truly singular, as the split is harmless for non-singular integrals, costs a fixed amount of work per observer that scales more weakly than the $O(N^2)$ kernel evaluation and $O(N^3)$ solve, and better resolves the sharply peaked integrands occurring for close conducting walls. We have observed consistent results with this method for conducting walls with a plasma-wall gap greater than approximately one cell of the coarser grid direction, and more specialized routines exist to enable closer gaps \cite{ying2006}.

When the plasma and wall are well separated, the cross-surface integrands are smooth and the wall collocation points can be chosen independently to best resolve the geometry. For close walls, however, the cross-surface singular patch must be centered on the source point closest to the observer to resolve the near-singular integrand, and walls parameterized independently of the plasma lose accuracy in this regime \cite{chance1997}. The wall collocation points must therefore be placed at the normal projections of the plasma collocation points onto the wall, so that each wall index $i$ is the closest wall point to plasma index $i$, which is valid provided the gap is small compared with the curvature scales of both surfaces. These are the same projections required for the near-singular scheme of \citet{ying2006}.

This singularity correction is defined in index space rather than physical space, a computationally efficient choice making the correction parameters dependent only on the index offset from the observer and allowing them to be assembled once for all observers. The singularity itself is physical, however, which imposes implicit requirements on the collocation grid. The polar transformation cancels the singularity for any grid choice, but the remaining integrand still depends on the polar angle $\omega$ if the grid is anisotropic. The dominant departure from isotropy is set by the cell aspect ratio $A = h_{\zeta}/h_{\theta}$, where $h_{\theta}$ and $h_{\zeta}$ are the physical grid spacings in the poloidal and toroidal directions. The support of $\eta$ is a disk in index space, but its image on the boundary (dashed orange in \cref{fig:singular_quadrature}(a)) is approximately an ellipse of axis ratio $A$. The angular trapezoidal rule is exact to leading order for $A=1$, but more generally it converges geometrically as $e^{-N_{\omega}/A}$ for $A \gg 1$, so the convergence rate degrades with increasing cell aspect ratio \cite{trefethen2014}. We therefore maintain accuracy by choosing the ratio $N_{\zeta}/N_{\theta}$ to hold $A$ of order unity in \cref{fig:singular_quadrature}(a) and throughout \cref{sec:implementation-results}. While we target the mean $A$ throughout this paper, we note that $A$ is a local quantity as the surface metric varies over the boundary, so the collocation points pack in strongly shaped regions with the quadrature error set by the largest local anisotropy. We reserve discussion of potential improvements to this formulation for \cref{sec:conclusion}.

\subsection{Singular Correction in the Matrix Formulation}
\label{subsec:matrix_sing_quad}
In order to preserve the matrix formulation desired for our vacuum response calculations, we must cast the singular treatment from \cref{subsec:malhotra_quadrature} into the collocation weights $w_{ij}$ from \cref{eq:discrete_fredholm} that allow us to write the discrete boundary integral equation in the form of \cref{eq:discrete_fredholm_matrix}. We will show that the single and double-layer potential operators can be written as
\begin{equation}
\label{eq:operator_components}
    \begin{split}
    \mathbb{S} &= \frac{4\pi^2}{N} \left( \mathbb{G} - \mathbb{E} \odot \mathbb{G} + \delta \mathbb{G}_{\mathrm{sing}} \right), \\
    \mathbb{D} &= \frac{4\pi^2}{N} \left( \mathbb{K} - \mathbb{E} \odot \mathbb{K} + \delta \mathbb{K}_{\mathrm{sing}} \right),
    \end{split}
\end{equation}
where $[\mathbb{G}]_{ij}=G(\mathbf{r}_i,\mathbf{r}_j)$,  $[\mathbb{K}]_{ij}=K(\mathbf{r}_i,\mathbf{r}_j)$, $[\mathbb{E}]_{ij}=\eta(\mathbf{r}_i,\mathbf{r}_j)$, $\delta \mathbb{G}_{\mathrm{sing}}$ and $\delta \mathbb{K}_{\mathrm{sing}}$ are correction terms from the singular integral in polar coordinates, and $\odot$ denotes the Hadamard (element-wise) product. The diagonal entries of $\mathbb{G}$ and $\mathbb{K}$ are set to zero to avoid division by zero in the kernels, and therefore the diagonals of $\mathbb{D}$ and $\mathbb{S}$ are entirely due to the singular correction. This can also be interpreted as $\eta_{ii} = 1$ by definition, so the first two terms cancel on the diagonal.
Because the smooth integral of \cref{eq:smooth_discrete} is evaluated entirely on the collocation points, it naturally lends itself to the matrix form with a weight of $(4\pi^2/N)(1 - \eta_{ij})$ and manifests as the first two terms in \cref{eq:operator_components}.
The singular integral, however, is evaluated off-grid. It can nonetheless be written as weights on collocation values by applying the transpose of the interpolation operator that places the polar nodes, as described in \citet{malhotra2019}.

Preserving the notation of \cref{subsec:malhotra_quadrature}, the singular integral for observer point $i$ is discretized as
\begin{equation}
\label{eq:discrete_sing_integral}
\begin{split}
    I^H_{i,\mathrm{sing}} &\approx \frac{4\pi^2}{N}  \frac{2\pi} {N_\omega} R_\eta^2\sum_{r=1}^{N_\rho} w^{\mathrm{GL}}_r \eta(\rho_r)  \rho_r\sum_{s=1}^{N_\omega}  H(\mathbf{r}_i, \mathbf{r}_{i,rs})f(\mathbf{r}_{i,rs}) \\
    &= \frac{4\pi^2}{N} \sum_{r=1}^{N_\rho} \lambda_r \sum_{s=1}^{N_\omega}  H(\mathbf{r}_i, \mathbf{r}_{i,rs})f(\mathbf{r}_{i,rs}),
\end{split}
\end{equation}
where $N_\rho$ is the order of the Gauss--Legendre quadrature and the number of radial points, $N_\omega$ is the number of angular quadrature points, $w^{\mathrm{GL}}$ are the Gauss--Legendre weights, and we have used the subscript $i,rs$ to differentiate between the polar nodes of each observer point. The quadrature weight, partition of unity value, and Jacobian term depend only on the radial index $r$ and are identical for every observer point, and so we have combined them into a single weight parameter
\begin{equation}
   \lambda_r = \frac{2\pi}{N_\omega } R_\eta^2 w^{\mathrm{GL}}_r \eta(\rho_r) \rho_r,
\end{equation}
which we have plotted in \cref{fig:singular_quadrature}(b). This combined weight vanishes at both ends of the radial integration domain, with $\lambda$ forced to $0$ by the partition of unity at $\rho=1$ and the polar Jacobian at $\rho=0$.

We use interpolation to obtain the geometry at the off-grid polar nodes. We denote by $\mathbb{P}$ the standard tensor-product Lagrange interpolation matrix of order $\ell$, whose row $(rs)$ carries the interpolation weights of the $\ell\times\ell$ patch nodes surrounding polar node $(r,s)$. It is an $N_\rho N_\omega \times D_\eta^2$ sparse matrix with exactly $\ell^2$ nonzeros per row. Like the polar quadrature weights, $\mathbb{P}$ is identical for every observer. The polar node geometry can then be obtained by multiplying a vector of the geometry values at each on-grid singular patch point for observer $i$ with $\mathbb{P}$, visualized in \cref{fig:singular_quadrature}(c) for a sample polar node. Following \citet{malhotra2019}, we interpolate the tangent vectors and then recompute the normal vector at each quadrature node for higher accuracy than directly interpolating the normal vector. Using the interpolated geometry, we can now evaluate the kernels at the polar nodes surrounding observer $i$
and collect them and their corresponding weights into vectors $\cvec{h}_i$ of length $N_\rho N_\omega$,
\begin{equation}
\label{eq:polar_samples}
\big[\cvec{h}_i\big]_{rs} = \lambda_r H\!\big(\mathbf{r}_i,\mathbf{r}_{i,rs}\big).
\end{equation}

The values of $f$ at the polar nodes are not independent data, and are instead supplied by the same operator $\mathbb{P}$ that generated the polar geometry, $f(\mathbf{r}_{i,rs}) = [\mathbb{P}\,\cvec{f}_i]_{rs}$, with $\cvec{f}_i$ the values of $f$ at the collocation points for the singular patch about observer $i$. The singular part of the boundary integrals then collapses \cref{eq:discrete_sing_integral} onto strictly on-grid contributions,
\begin{equation}
    I^H_{i,\mathrm{sing}} \approx \frac{4\pi^2}{N}\sum_{r,s}\big[\cvec{h}_i\big]_{rs}\big[\mathbb{P}\,\cvec{f}_i\big]_{rs} = \frac{4\pi^2}{N}\big(\mathbb{P}^{\mathsf T}\cvec{h}_i\big)\cdot\cvec{f}_i,
\end{equation}
so $f$ is never required off-grid and we retain the desired matrix form. In the representation on the right-hand side, the singular integral becomes a set of weights acting on collocation values, with $\mathbb{P}^{\mathsf T}\cvec{h}_i$ precisely the $D_\eta^2$ nonzero entries in row $i$ of $\delta \mathbb{G}_\mathrm{sing}$ and $\delta \mathbb{K}_\mathrm{sing}$ for $H = G$ and $H=K$, respectively, when the patch points are mapped back into their respective global source column indices. The total singular correction is then assembled row by row for each observer. This adjoint application of $\mathbb{P}$, contrasted with its forward application between \cref{fig:singular_quadrature}(c) and (d), follows \citet{malhotra2019}, where the weights $\mathbb{P}^{\mathsf T}\cvec{h}_i$ are precomputed once per observer and reused for every surface density. Here, we instead assemble them into the rows of the dense operators $\mathbb{D}$ and $\mathbb{S}$, including the cross-surface plasma--wall blocks, which allows the correction to enter the matrix formulation and the symmetry reduction of \cref{sec:field-period-symmetry}.

With \cref{eq:operator_components}, the operators $\mathbb{D}$ and $\mathbb{S}$ of \cref{eq:discrete_fredholm_matrix} are fully defined, and the response matrix, vacuum energy, and surface inductance follow. We will now show in \cref{sec:field-period-symmetry} how discrete symmetries reduce the cost of forming and inverting them.

\section{Field-Period Symmetry}
\label{sec:field-period-symmetry}

While stellarators are non-axisymmetric, they are frequently designed to include other discrete symmetries, most notably field-period symmetry. A stellarator with $n_{\mathrm{fp}}$ identical field periods is invariant under a rigid rotation by $2\pi/n_{\mathrm{fp}}$ in the toroidal direction, leading to the well-known decoupled mode families in Fourier space \cite{nuhrenberg1996a}. In this section, we present how field-period symmetry can be leveraged in the boundary integral formulation to reduce the storage required by a factor of $n_{\mathrm{fp}}$ and the cost of the linear solve by a factor of $n_{\mathrm{fp}}^2$; the trade-off between kernel evaluations and operator storage is discussed in \cref{sec:implementation-results}. In the following, we assume that the wall and plasma surfaces share the same field-period symmetry.

When the boundary $\Gamma$ of the vacuum region possesses an $n_{\mathrm{fp}}$-fold symmetry, an observer and a source interact through a kernel that depends only on the toroidal period offset $d$ between them, taken modulo $n_{\mathrm{fp}}$. The boundary operators in \cref{eq:discrete_fredholm_matrix} are therefore block-circulant matrices with blocks indexed by $d$
\begin{equation}
\label{eq:block_circulant}
\mathbb{D} =
\begin{pmatrix}
\mathbb{D}_0 & \mathbb{D}_{1} & \cdots & \mathbb{D}_{n_{\mathrm{fp}}-1} \\
\mathbb{D}_{n_{\mathrm{fp}}-1} & \mathbb{D}_0 & \cdots & \mathbb{D}_{n_{\mathrm{fp}}-2} \\
\vdots & & \ddots & \vdots \\
\mathbb{D}_{1} & \mathbb{D}_{2} & \cdots & \mathbb{D}_0
\end{pmatrix},
\end{equation}
and likewise for $\mathbb{S}$. We define $\mathbb{D}_0$ to include the $\tfrac{1}{2}\mathbb{I}_{N_{\mathrm{per}}}$ term of \cref{eq:discrete_fredholm_matrix}, which preserves the block-circulant structure and resides entirely in the diagonal blocks. Each block is an $N_{\mathrm{per}} \times N_{\mathrm{per}}$ matrix, where $N_{\mathrm{per}} = N/n_{\mathrm{fp}}$ is the number of collocation points per period. These matrices are fully specified by the first row of blocks $\{\mathbb{D}_d, \mathbb{S}_d\}_{d=0}^{n_{\mathrm{fp}}-1}$, for a total of $N_{\mathrm{per}} \times N$ entries, or $2N_{\mathrm{per}} \times 2N$ for $\mathbb{D}$ and $2N_{\mathrm{per}} \times N$ for $\mathbb{S}$ with a conducting wall. This matrix property occurs because the Green's function depends only on the observer--source distance, so rotating both by one field period leaves the kernel unchanged and observers in different periods see the same sources up to a cyclic relabeling; however, the sources must still span the entire boundary to capture inter-period coupling. This reduction does not affect our singular correction scheme in \cref{sec:singular}, as the correction is applied row-wise in each block for the observers.

A block-circulant matrix can be block-diagonalized by the discrete Fourier transform over the period index, reducing $\mathbb{D}$ and $\mathbb{S}$ to $n_{\mathrm{fp}}$ diagonal blocks $\hat{\mathbb{D}}_k$ and $\hat{\mathbb{S}}_k$. Each block represents the $k$th mode family, containing the modes with identical $k = n \bmod n_{\mathrm{fp}}$, for a total of $n_{\mathrm{fp}}$ decoupled families, or $1 + \lfloor n_{\mathrm{fp}}/2 \rfloor$ distinct families for a real perturbation, since $k$ and $n_{\mathrm{fp}} - k$ are complex conjugates \cite{nuhrenberg1996a}. It can be shown that these blocks are given by \cite{olson2014}
\begin{equation}
    \label{eq:residue_blocks}
    \hat{\mathbb{D}}_k = \sum_{d=0}^{n_{\mathrm{fp}}-1} \mathbb{D}_d\,e^{-2\pi i \frac{k}{n_{\mathrm{fp}}}d},
    \qquad
    \hat{\mathbb{S}}_k = \sum_{d=0}^{n_{\mathrm{fp}}-1} \mathbb{S}_d\,e^{-2\pi i \frac{k}{n_{\mathrm{fp}}}d}
    \end{equation}
for each $k$. In this basis \cref{eq:discrete_fredholm_matrix} decouples into $n_{\mathrm{fp}}$
independent systems of size $N_{\mathrm{per}}$, or $2N_{\mathrm{per}}$ when a conducting wall is present,
\begin{equation}
\label{eq:residue_solve}
\hat{\mathbb{D}}_k\,\cvec{\varphi}_k = -\hat{\mathbb{S}}_k\,\cvec{b}_k ,
\end{equation}
where the vectors $\cvec{\varphi}_k$ and $\cvec{b}_k$ denote the $k$th mode family component of $\cvec{\varphi}$ and $\cvec{b}$, respectively, transformed in a similar way to \cref{eq:residue_blocks}.

For each family, we form the reduced operator $\hat{\mathbb{D}}_k^{-1}\hat{\mathbb{S}}_k$, project it onto the subset of modes with $n \bmod n_{\mathrm{fp}} = k$ using the single-period Fourier basis $\mathbb{F}_k$, and scatter the result into the corresponding diagonal block of $\mathbb{W}_V$,
\begin{equation}
\label{eq:Wv_class}
\mathbb{W}_V^{(k)} = \frac{\chi'^2}{2\mu_0}\frac{4\pi^2}{N_{\mathrm{per}}}\,
\mathbb{Q}_k\, \mathbb{F}_k^\dagger \,
\big(\hat{\mathbb{D}}_k^{-1}\hat{\mathbb{S}}_k\big)\big|_{\Gamma_p}\,
\mathbb{F}_k\,\mathbb{Q}_k ,
\end{equation}
where $\mathbb{F}_k$ and $\mathbb{Q}_k$ are the restrictions of $\mathbb{F}$ and $\mathbb{Q}$ to the modes of family $k$, and $\mathbb{F}_k$ is evaluated on the $N_{\mathrm{per}}$ collocation points of a single period.

This method reduces both computational cost and storage. Because the operators are fully determined by their first block row, they contain only $N_{\mathrm{per}} \times N$ distinct entries rather than $N \times N$, and the dense matrix inversion changes from one $O(N^3)$ solve to $n_{\mathrm{fp}}$ solves of size $N/n_{\mathrm{fp}}$, reducing the cost by a factor of $n_{\mathrm{fp}}^2$. For $n_{\mathrm{fp}}=1$ the decomposition only adds overhead, and we fall back to the direct solve of \cref{sec:response_matrix}.

In addition to field-period symmetry, we exploit two further properties of the boundary operators. First, the reality of the boundary operators requires that $\hat{\mathbb{D}}_{n_{\mathrm{fp}}-k} = \hat{\mathbb{D}}^*_k$ and $\hat{\mathbb{S}}_{n_{\mathrm{fp}}-k} = \hat{\mathbb{S}}_k^*$. A family and its conjugate can therefore share a single assembly and factorization, reducing the number of operators formed and factored to the $1 + \lfloor n_{\mathrm{fp}}/2 \rfloor$ distinct families counted above. Second, most stellarator equilibria are designed to be stellarator symmetric, with the boundary invariant under $R(-\theta,-\zeta) = R(\theta,\zeta)$, $Z(-\theta,-\zeta) = -Z(\theta,\zeta)$. This reflection pairs each collocation point with its mirror image, and recombining each pair into its sum and difference makes every reduced operator real, allowing for simplified arithmetic. Because this property maps $n$ to $-n$, the self-conjugate families $k=0$ and, for even $n_{\mathrm{fp}}$, $k = n_{\mathrm{fp}}/2$ are mapped to themselves and further decouple into even and odd parity problems of half size. Together, these properties halve the kernel evaluations and storage and reduce the factorization cost.

\section{Numerical Implementation and Results}
\label{sec:implementation-results}

We have implemented the preceding formulation in the new Julia GPEC version 2.0 \cite{logan2026} openly available at \url{https://github.com/OpenFUSIONToolkit/GPEC}. The code accepts the plasma boundary in Cartesian coordinates as input, equally spaced in the straight-field-line angles in one field period, together with the desired Fourier modes. The boundary is then expanded to the full torus and we compute the surface position, tangent, and normal vectors using periodic bicubic splines on the collocation grid points. We detect the sign of $\sigma$ in the normal vector automatically based on the handedness of the surface parameterization, ensuring it points out of the vacuum region as required by the exterior Neumann problem in \cref{sec:response_matrix}. The code currently supports a conformal conducting wall, offset in the normal direction by a user-supplied fraction of the plasma minor radius and built at the same resolution as the plasma boundary. Each wall point is then the normal projection of its plasma point, satisfying the requirement of \cref{subsec:malhotra_quadrature} by construction. Such offsets can self-intersect once the gap exceeds the concave radius of curvature of the plasma boundary \cite{fu2026a}, so the code rejects any offset which folds the surface. We then assemble and solve the exterior system for the vacuum response matrix, and only form the interior system if the inductance matrix is requested to avoid the unnecessary additional factorization.

The cost for the 3D boundary solve at realistic resolutions is dominated by the $O(N^3)$ matrix factorization required to invert the linear system, differing from axisymmetric cases where the smaller system sizes make the kernel build equal to or even greater than the factorization cost. Even more limiting is the $O(N^2)$ memory required to store the dense boundary operators, quickly reaching several GB at $N \gtrsim 10^4$. With this ordering in mind, we have made several choices to make the problem more numerically tractable. First, rather than assembling and storing the entire $N_{\mathrm{per}} \times N$ block row of \cref{eq:block_circulant} and then forming each decoupled mode-family block of \cref{eq:residue_blocks}, we combine the Green's function evaluation with the multiplication by the phase $\exp\left(-2\pi i k d / n_{\mathrm{fp}}\right)$, accumulating directly into $\hat{\mathbb{D}}_k$ and $\hat{\mathbb{S}}_k$. This method requires only one $N_{\mathrm{per}} \times N_{\mathrm{per}}$ block to be formed at a time, improving the storage reduction to $n_{\mathrm{fp}}^2$ from $n_{\mathrm{fp}}$ at the cost of re-evaluating the kernel for each mode family. Second, similar to the axisymmetric code \cite{chance1997}, we avoid forming $\mathbb{R}$ explicitly by first multiplying the Fourier operator onto $\hat{\mathbb{S}}_k$, reducing the number of triangular solves after factorization from $N_{\mathrm{per}}$ down to the $M_k$ modes retained in family $k$. Because the exterior and interior operators required for the surface inductance calculation differ only by a shift of the diagonal, the code currently assembles the matrices once and stores two factorizations, with and without the shift.

The singular quadrature of \cref{sec:singular} is a Julia translation of the C++ implementation in the BIEST code \cite{biest}, validated against it by comparing the discretized single- and double-layer operators and integrated into our formulation. The scheme is defined by the patch size $D_\eta$, radial and angular quadrature orders $N_\rho$ and $N_\omega$, and the Lagrange interpolation order $\ell$. Previous works have scaled the patch size with the total grid size \cite{ying2006, malhotra2019} to balance convergence with computational cost. At the grid sizes in our benchmarks, we have found this scaling has little impact on our error and we therefore fix $D_\eta = 23$, $N_\rho = 20$, $N_\omega = 2N_\rho = 40$, and $\ell = 5$ in this work. The parameters $\lambda$, $\eta$, and $\mathbb{P}$ are pre-assembled before matrix assembly, with $\mathbb{P}$ stored as a sparse matrix for rapid evaluation. Because each observer requires a $D_\eta \times D_\eta$ patch of distinct source points, we require that $\min \left(N_\theta, N_\zeta \right) \geq D_\eta$.

The code is written entirely in Julia and runs at the resolutions used here on a laptop. We have parallelized the kernel evaluation over observer points, which is a natural choice due to the row-wise assembly and singular treatment described above, and keeps the per-thread memory requirements low compared to a parallelization over mode families. All runtimes reported below use $6$ threads for both the observer loop and the factorization with OpenBLAS. We also select the operator element type per mode family. The self-conjugate families with $k = 0$ and $k = n_{\mathrm{fp}}/2$ (for even $n_{\mathrm{fp}}$) have real phases $e^{-2\pi i k d/n_{\mathrm{fp}}} = \pm 1$ and therefore real boundary operators, which we store and factor in real arithmetic to reduce cost. The remaining families are complex unless the boundary is detected to be stellarator symmetric, in which case the symmetry-adapted basis of \cref{sec:field-period-symmetry} renders every family real. The peak operator storage is unchanged when any family is complex and halved when every family is real, as it is for $n_{\mathrm{fp}} \le 2$ or for any stellarator-symmetric boundary.

\subsection{Axisymmetric Benchmarking}

First, we compare the 3D version of the VACUUM code to its axisymmetric counterpart. Although the axisymmetric and non-axisymmetric codes use the same fundamental approach and are separate branches in the same Julia code, the axisymmetric code represents the boundary as a 2D contour and uses toroidal Green's functions with a leading-order singularity correction scheme \cite{chance1997}, while the 3D code uses the full toroidal discretization and high-order correction from \cref{sec:singular}. Agreement between them therefore demonstrates proper implementation of both the high-order quadrature and the non-axisymmetric extension.

\begin{figure}[]
    \centering
    \includegraphics[width=\columnwidth]{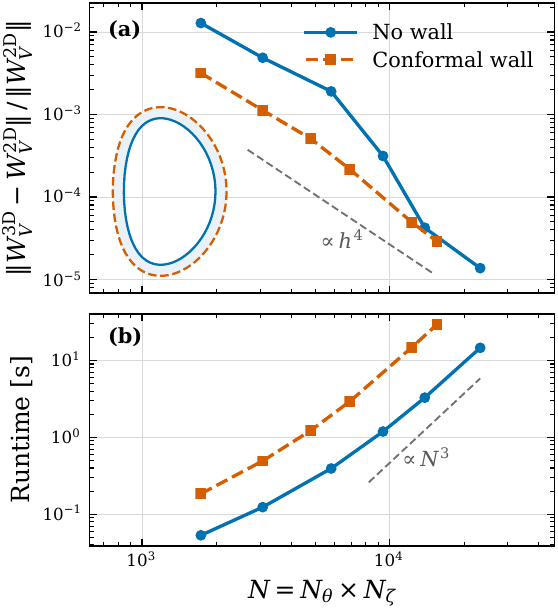}
    \caption{(a) Frobenius norm of the difference between the 3D and 2D response matrices normalized by the 2D norm for an analytic Solov'ev equilibrium with and without a conformal wall in solid blue and dashed orange, respectively. The 2D contours of the boundaries for the plasma and conformal wall are shown as an inset in their corresponding line styles. The response matrix includes $n=1$ and $m\in\left[ -15, 15 \right]$, and the 2D reference case was computed with $N_\theta = 2048$. We fix $N_\zeta = 4 N_\theta / 3$ across the scan and show the asymptotic scaling proportional to $h^4$, where $h$ is the grid spacing. (b) Runtimes as measured on a 2023 MacBook M3 Pro with $6$ threads for the cases in (a).
    \label{fig:2D_vs_3D_vacuum}}
\end{figure}

We use an analytic Solov'ev tokamak equilibrium with a major radius of $1\,\mathrm{m}$, minor radius of $0.33\,\mathrm{m}$, elongation of $1.6$, and on-axis safety factor and toroidal field of $1.9$ and $1\,\mathrm{T}$, respectively. We compute the vacuum response matrix with $n=1$ and $m \in \left[ -15, 15 \right]$ for a range of grid resolutions, both with and without a conformal wall offset by $8.5\,\mathrm{cm}$ from the plasma boundary in \cref{fig:2D_vs_3D_vacuum}. In \cref{fig:2D_vs_3D_vacuum}(a), we plot the relative error of the 3D vacuum response matrix with respect to a 2D high-resolution reference case with $N_\theta = 2048$, while \cref{fig:2D_vs_3D_vacuum}(b) shows the code runtimes as measured on a 2023 MacBook M3 Pro. We fix $N_\zeta = 4 N_\theta / 3$ in the scan, holding the grid cell aspect ratio constant. The minimum grid resolution was determined by the Nyquist limit and the maximum based on the available RAM on the laptop.

The relative error falls to $O(10^{-5})$ for both the no-wall and conformal-wall cases at the highest resolution scan point. On the same subplot, we have included a scaling line proportional to $h^4$ illustrating the approximate convergence of our method both with and without a conducting wall, where $h$ is the grid spacing and proportional to $N^{-1/2}$. The error is pre-asymptotic at these resolutions, with contributions that can change sign across the scan and produce departures from the general $h^4$ trend such as the steepening of the no-wall curve near $N \approx 10^4$. These convergence properties are only meaningful while the difference remains well above the discretization error of the 2D reference itself, which we estimate at $O(10^{-6})$ from the self-convergence of the 2D code at $N_\theta = 2048$. The exponential convergence of the high-order scheme has been demonstrated against an analytic reference by \citet{malhotra2019a}. The convergence of the similar approach by Merkel \cite{merkel1986} has been measured at $h^2$ in the same analytic test \cite{malhotra2019a}. The practical consequence of this high-order convergence is a reduction in the grid resolution required for a given error target compared to leading-order methods.

Because $\delta W_V$ is a strictly real quantity, the vacuum response matrix $\mathbb{W}_V$ must be a Hermitian matrix and any non-Hermitian components are due to discretization error. In the axisymmetric comparison, we also computed the non-Hermitian residual $\|\mathbb{W}_V - \mathbb{W}_V^\dagger\| / \|\mathbb{W}_V + \mathbb{W}_V^\dagger\|$, and found that it falls at approximately $h^4$, tracking the measured convergence rate of the solve though not its magnitude. As a result, we use this non-Hermitian residual as an error proxy in the code to verify a converged result without a reference point, similar to the axisymmetric code \cite{chance1997}.

In \cref{fig:2D_vs_3D_vacuum}(b), we observe that as the grid discretization grows, the runtime approaches the $N^3$ scaling expected in factorization-dominated calculations; however, all cases still finish in under $30\,\mathrm{s}$. Including a conducting wall doubles the number of collocation points, increasing runtime by a factor of $4$ to $8$ depending on whether kernel evaluation or factorization dominates, and lowers the maximum resolution reachable on a given architecture through the added memory requirements.
 
\subsection{Non-axisymmetric Benchmarking}

\begin{figure}[]
    \centering
    \includegraphics[width=\columnwidth]{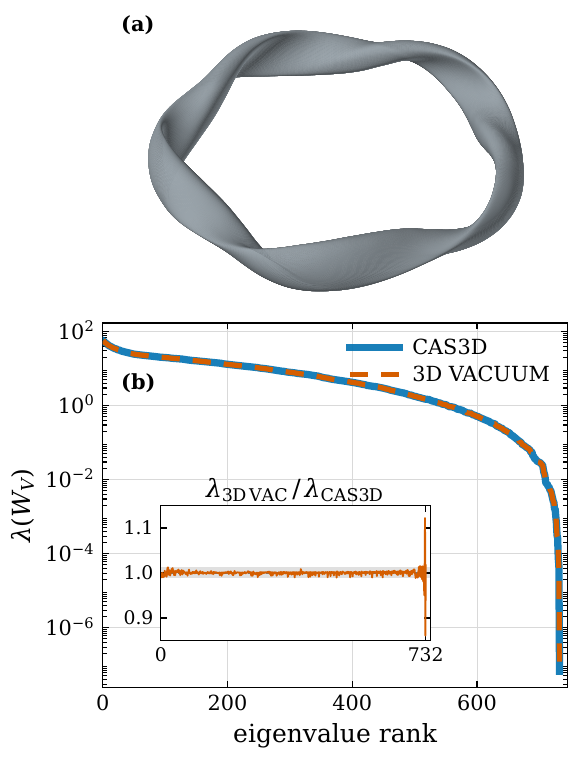}
    \caption{(a) W7-X boundary with $n_{\mathrm{fp}} = 5$ used for this code comparison. (b) Eigenvalues $\lambda(\mathbb{W}_V)$ of the $k=1$ mode family vacuum response matrix, sorted in descending rank, for CAS3D in solid blue and 3D VACUUM in dashed orange. We retain poloidal and toroidal harmonics up to $30$, giving $732$ modes in the CAS3D $m \geq 0$ sine/cosine basis, of which we plot the symmetrized cosine block; the sine block is isospectral. The grid resolutions are $N_\theta = 180, N_\zeta = 900$ for CAS3D and $N_\theta = 160, N_\zeta = 400$ for 3D VACUUM. The inset shows the ratio $\lambda_{\mathrm{3D\,VAC}}/\lambda_{\mathrm{CAS3D}}$ at each rank on a linear scale spanning $\pm 15\%$, with the $\pm 1\%$ band shaded in grey.
    \label{fig:CAS3D_benchmark}}
\end{figure}

To verify our implementation in non-axisymmetric geometries, we benchmark the 3D VACUUM calculation against the CAS3D \cite{nuhrenberg2021} vacuum energy calculation in the no-wall limit, which employs the leading-order singularity correction scheme \cite{merkel1986, merkel1996}. We use an $n_{\mathrm{fp}} = 5$ W7-X equilibrium in standard configuration \cite{sunnpedersen2015,panici2023}, visualized in \cref{fig:CAS3D_benchmark}(a), with Fourier harmonics in Boozer coordinates up to $30$ in both the toroidal and poloidal directions for the vacuum calculation. We use relatively high-resolution grids of $N_\theta = 180, N_\zeta = 900$ and $N_\theta = 160, N_\zeta = 400$ for CAS3D and 3D VACUUM, respectively, to ensure convergence. We transform the 3D VACUUM matrix into the CAS3D cosine/sine convention and matrix normalization \cite{schwab1993, merkel1996} and plot the eigenspectra of each output for the $k=1$ mode family in \cref{fig:CAS3D_benchmark}(b).

The eigenspectra of the two codes are nearly indistinguishable across the $732$ modes of the $k=1$ mode family and nearly nine decades of eigenvalue magnitude (with similar findings for the $k=0$ and $k=2$ mode families). In the inset, we plot the ratio of each eigenvalue, shading the $\pm 1\%$ band in light grey. Almost all eigenvalues agree to within this $1\%$ margin, and only the smallest near-resonant modes with $m - nq \approx 0$ show a discrepancy of up to $14\%$. These modes are suppressed by the factor $(m-nq)^2$ carried by $\mathbb{Q}$ in \cref{eq:Wv_fourier}, placing them decades of magnitude below the largest eigenvalue where the relative comparison is set by the absolute error floor of each code. The full matrices also agree well, with a relative Frobenius error between the two of $\approx 2\%$.

\begin{figure}[]
    \centering
    \includegraphics[width=\columnwidth]{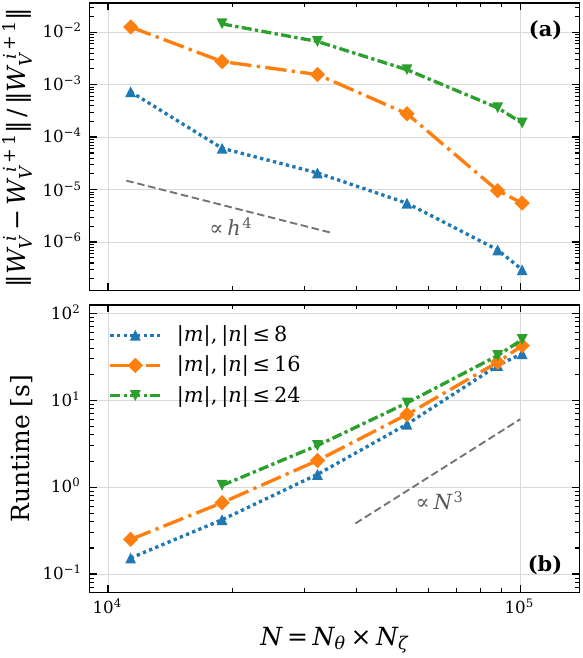}
    \caption{(a) Self-convergence of the $k=1$ mode family vacuum response matrix $\mathbb{W}_V$ versus grid resolution $N=N_\theta\times N_\zeta$ for the same $n_{\mathrm{fp}} = 5$ W7-X equilibrium as \cref{fig:CAS3D_benchmark}. We show three poloidal/toroidal mode truncations and only plot runs with sufficient grid points to satisfy the Nyquist criterion. We fix $N_\zeta \approx 6.5 N_\theta$ for a cell aspect ratio $A \approx 1.5$, and the dashed line again shows $h^4$ for reference.  (b) Runtimes as measured on a 2023 MacBook M3 Pro with $6$ threads for the cases in (a), with the dashed line again showing the $N^3$ scaling. Runtime for the finest grid is omitted from (b) at all three truncations, where memory compression on the test machine inflates the wall time; the response matrix is unaffected and is retained in (a).}
    \label{fig:nonaxisym_scan}
\end{figure}

To extend the findings of \cref{fig:2D_vs_3D_vacuum} to non-axisymmetric geometry, we plot the self-convergence of the 3D VACUUM code along with its runtime for increasing grid resolution in \cref{fig:nonaxisym_scan}. In \cref{fig:nonaxisym_scan}(a), we plot the normalized response matrix difference between successive resolutions and in \cref{fig:nonaxisym_scan}(b) we again plot the runtime. Because toroidal harmonics no longer decouple in non-axisymmetric geometry, each solve must retain the full set of $(m,n)$ pairs within a mode family rather than the single $n$ of an axisymmetric run, increasing both the runtime and the error at a given grid. To show this effect, we plot runs retaining Fourier harmonics up to $8$ in blue, $16$ in orange, and $24$ in green. The finest grid reaches $\approx 7.5\,\mathrm{GB}$ of peak memory.

The error is set by the larger of two contributions: the surface discretization and the accuracy of the smooth and singular quadratures. Because we hold $D_\eta$, $N_\rho$, $N_\omega$, and $\ell$ fixed and refine only $N$, only the former improves across the scan and we would expect the error to eventually saturate at the level set by the quadrature parameters. We do not reach that saturation here, as the differences are still falling at the finest grid for all three truncations. The Fourier truncation sets the highest harmonics the grid must resolve; increasing it from $8$ to $24$ at fixed grid resolution raises the error by more than two orders of magnitude, highlighting the expected result that finer grids are required to resolve higher harmonics to a target error level. The convergence rate approximately agrees with the $h^4$ of \cref{fig:2D_vs_3D_vacuum} at coarser grid sizes but rises toward the finest grids, with a least-squares fit over the entire scan giving $h^{6.6}$, $h^{7.0}$, and $h^{5.2}$ for the $8, 16$, and $24$ maximum mode number cases, respectively. The steepening slope in our measurements is the expected signature of the exponential convergence of this scheme \cite{malhotra2019a}.

Even at $N \approx 10^5$ points on the full torus, the runtimes remain under one minute and peak memory stays below $8\,\mathrm{GB}$, comparable to the axisymmetric runs of \cref{fig:2D_vs_3D_vacuum} at far smaller $N$ due to our symmetry-related reductions. The number of included modes, varying from $\approx 100$ to $\approx 1000$ here, enters only through the Fourier projection and triangular solves, so its effect on runtime is secondary to the grid size, especially at higher resolutions when the $N^3$ factorization dominates. At $N \approx 1.9 \times 10^4$ the runtime rises from $0.42\,\mathrm{s}$ to $1.05\,\mathrm{s}$ across a factor of $10$ in mode count, and at $N \approx 10^5$ from $35\,\mathrm{s}$ to $50\,\mathrm{s}$. Because the grid discretization sets the memory requirements, additional modes affect solve time but not peak memory.

\section{Conclusion} 
\label{sec:conclusion}
We have extended the vacuum energy calculations in the VACUUM code \cite{chance1997} to non-axisymmetric geometries, freely available through the open-source GPEC (\url{https://github.com/OpenFUSIONToolkit/GPEC}), soon to be released as version 2.0. This new  boundary integral equation solver uses a high-order singularity correction scheme \cite{malhotra2019, malhotra2019a} and leverages the block-circulant matrix structure under field-period symmetry to improve performance. The theory applies to plasmas both with and without a nearby conducting wall.
We have benchmarked the implementation against the axisymmetric VACUUM code for a Solov'ev tokamak, both with and without a conducting wall, and against the CAS3D code for W7-X, with runtimes of tens of seconds at up to $10^5$ grid points on a laptop.

While our numerical implementation enables moderate grid sizes, several techniques common in boundary integral equation solvers, such as fast multipole methods or matrix-free solvers, could provide higher performance \cite{malhotra2019a}. Furthermore, the singular quadrature scheme here is built for on-surface integrals and breaks down for conducting walls extremely close to the plasma, which could be addressed by implementing the near-singular methods of \citet{ying2006}; this feature would also enable evaluation of the vacuum fields just outside the surface. Our conformal wall offset in the normal direction can also create self-intersecting surfaces for distant walls in strongly shaped stellarator geometries, and supporting smooth formulations of offset surfaces \cite{fu2026a} in addition to user-supplied custom wall geometries could improve the flexibility of the code. 

Finally, as discussed in \cref{subsec:malhotra_quadrature}, the singular quadrature scheme is defined in index space, which makes $\eta$, $\lambda_r$, and $\mathbb{P}$ reusable for every observer at the cost of stricter requirements on the collocation grid due to the singularity being in physical space. Because the partition of unity spans an equal number of grid points in the poloidal and toroidal indices, we require comparable physical grid spacing in both directions to minimize the singular quadrature error; however, this typically results in $N_\zeta \gg N_\theta$ for the high-aspect-ratio boundaries common in stellarators even if not necessary to resolve the Fourier harmonics. Separate index widths $D_{\eta, \theta}$ and $D_{\eta,\zeta}$ could relax this constraint for uniformly anisotropic grids without the loss of computational efficiency, and a more general treatment could define the correction region using the local surface metric to make $\mathrm{supp}(\eta)$ a disk of fixed physical radius. This approach would accommodate the grid packing common in straight-field-line coordinates in strongly shaped regions, with an extreme example being near a magnetic X-point. This would come with the trade-off of making all singular quadrature parameters observer-dependent, and the merits of this implementation require further evaluation.

This development is the first step toward enabling several critical MHD calculations in stellarators. We are actively coupling the $\mathbb{W}_V$ calculation with a fixed boundary stability code \cite{glasser2025, jang2026a} to provide rapid finite mode number calculations of free boundary stellarator stability for equilibrium optimization, both with and without a conducting wall close to the plasma. We will then use the 3D surface inductance matrix to calculate perturbed stellarator equilibria \cite{parkComputationThreedimensionalTokamak2007a, nuhrenberg2012} and quantify the effect of externally applied fields. Because the boundary potential $\varphi$ is obtained as part of the same solve, the perturbed magnetic field can then be rapidly evaluated in the vacuum region for scrape-off layer or divertor calculations.

\section*{Acknowledgements}
The authors thank Dhairya Malhotra for making BIEST openly available and for comments on the attribution and presentation of the singular quadrature in this manuscript. This material is based upon work supported by the U.S. Department of Energy, Office of Science, Office of Advanced Scientific Computing Research, Department of Energy Computational Science Graduate Fellowship under Award Number DE-SC0024386. The work was also supported by the U.S. Department of Energy, Office of Science, Office of Fusion Energy Sciences under Award DE-SC0022272.

\section*{Declaration of generative AI and AI-assisted technologies in the manuscript preparation process}
During the preparation of this work the authors used Claude (Anthropic) to produce an initial Julia translation of the singular quadrature routines in the BIEST library \cite{biest} and the first versions of the field-period and stellarator symmetry reductions, and to assist with subsequent code implementation and debugging. The authors also used it to write plotting scripts that produce figures from existing data, verify completeness of references, and improve overall readability of the manuscript. After using this tool, the authors reviewed and edited all code, figures, and text as needed and take full responsibility for the content of the published article.

\section*{Data availability}
GPEC is available at \url{https://github.com/OpenFUSIONToolkit/GPEC}.
The equilibria and scripts used for benchmarking are available from the corresponding author on request.

\printcredits

\bibliographystyle{elsarticle-num-names}

\bibliography{bibliography}

\end{document}